\documentclass[preprint,3p,times,twocolumn]{elsarticle}

\usepackage[utf8]{inputenc}
\usepackage{natbib}
\usepackage{graphicx}
\usepackage{xcolor}
\usepackage{xspace}
\usepackage{mathtools}

\usepackage{hyperref}

\newcommand{\PbPb}         {\mbox{Pb--Pb}\xspace}

\usepackage{siunitx}
\DeclareSIUnit\clight{\text{\ensuremath{c}}}
\DeclareSIUnit\micron{\micro\metre}
\DeclareSIUnit\mrad{\milli\rad}
\DeclareSIUnit\gauss{G}
\DeclareSIUnit\eVperc{\eV\per\clight}
\DeclareSIUnit\nanobarn{\nano\barn}
\DeclareSIUnit\picobarn{\pico\barn}
\DeclareSIUnit\femtobarn{\femto\barn}
\DeclareSIUnit\attobarn{\atto\barn}
\DeclareSIUnit\zeptobarn{\zepto\barn}
\DeclareSIUnit\yoctobarn{\yocto\barn}
\DeclareSIUnit\nb{\nano\barn}
\DeclareSIUnit\pb{\pico\barn}
\DeclareSIUnit\fb{\femto\barn}
\DeclareSIUnit\ab{\atto\barn}
\DeclareSIUnit\zb{\zepto\barn}
\DeclareSIUnit\yb{\yocto\barn}

\newcommand{\nineH}        {$\sqrt{s}~=~0.9$~Te\kern-.1emV\xspace}
\newcommand{\seven}        {$\sqrt{s}~=~7$~Te\kern-.1emV\xspace}
\newcommand{\twoH}         {$\sqrt{s}~=~0.2$~Te\kern-.1emV\xspace}
\newcommand{\twosevensix}  {$\sqrt{s}~=~2.76$~Te\kern-.1emV\xspace}
\newcommand{\five}         {$\sqrt{s}~=~5.02$~Te\kern-.1emV\xspace}
\newcommand{\twosevensixnn}{$\sqrt{s_{\mathrm{NN}}}~=~2.76$~Te\kern-.1emV\xspace}
\newcommand{\fivenn}       {$\sqrt{s_{\mathrm{NN}}}~=~5.02$~Te\kern-.1emV\xspace}

\newcommand{\GeVc}         {Ge\kern-.1emV/$c$\xspace}
\newcommand{\MeVc}         {Me\kern-.1emV/$c$\xspace}
\newcommand{\TeV}          {Te\kern-.1emV\xspace}
\newcommand{\GeV}          {Ge\kern-.1emV\xspace}
\newcommand{\MeV}          {Me\kern-.1emV\xspace}
\newcommand{\GeVmass}      {Ge\kern-.2emV/$c^2$\xspace}
\newcommand{\MeVmass}      {Me\kern-.2emV/$c^2$\xspace}

\newcommand{\trento}{\protect\scalebox{1}{T$_{\text{R}}$ENT}o\xspace}

\def \FigPath {./Figures}

\begin{document}

\begin{frontmatter}
\title{$\Lambda$ polarization from thermalized jet energy}

\author[add1]{Willian Matioli Serenone\corref{cor1}}
\author[add1]{Jo\~{a}o Guilherme Prado Barbon}
\author[add1]{David Dobrigkeit Chinellato}
\author[add2]{Michael Annan Lisa}
\author[add3,add4]{Chun Shen}
\author[add1]{Jun Takahashi}
\author[add1]{Giorgio Torrieri}

\address[add1]{Instituto de F\'{i}sica Gleb Wataghin, Universidade Estadual de Campinas, Campinas, Brasil}
\address[add2]{The Ohio State University, Columbus, Ohio, USA}
\address[add3]{Department of Physics and Astronomy, Wayne State University, Detroit, MI 48201, USA}
\address[add4]{RIKEN BNL Research Center, Brookhaven National Laboratory Upton, NY 11973, USA}

\cortext[cor1]{Corresponding author}

\date{February 2021}
\begin{abstract}
    We examine the formation of vortical ``smoke rings'' as a result of thermalization of energy lost by a jet.
    We simulate the formation and evolution of these rings using hydrodynamics and define an observable that allows to probe this phenomenon experimentally.
    We argue that observation of vorticity associated with jets would be an experimental confirmation of the thermalization of the energy lost by quenched jets, and also a probe of shear viscosity.
\end{abstract}
\end{frontmatter}

\section{Introduction}

Two of the most studied results in heavy ion physics at ultra-relativistic energies are
jet energy loss~\cite{Gyulassy:1990ye, Wang:1991xy, Wiedemann:2009sh, Aad:2010bu, Chatrchyan:2011sx}
and fluid behavior~\cite{Schafer:2009dj, Werner:2012xh,  Heinz:2013th, Gale:2013da, Shuryak:2014zxa, Shen:2020mgh}. The first 
shows that colored degrees of freedom form ``a medium'' 
opaque to fast partons, and the second shows this medium 
thermalizes very quickly and subsequent evolution is nearly inviscid. Both results are 
usually interpreted as evidence that the medium created in heavy ion collisions is 
a ``strongly coupled liquid''.

However, considerable theoretical uncertainty exists
regarding the fate of the energy lost by the jet. 
If the plasma is a very good fluid it is a reasonable hypothesis that
the jet energy should thermalize and contribute to the fluid flow gradients.
However, we do not have a clear experimental signature
of this. Partially, this is because the models of parton-medium interaction are inconclusive \cite{Blaizot:2015lma}, and partially it is
because direct signatures of fluid behavior, such as ``Conical flow'',
have not been conclusively observed \cite{Takahashi:2009na, Alver:2010gr}.

Recently, a new intriguing manifestation of hydrodynamic behavior has been 
found: $\Lambda$ polarization, measurable via parity violating 
decays~\cite{Becattini:2020ngo}. It seems 
to be aligned to the global vorticity of the fluid and, to an extent, 
with near-ideal hydrodynamic vorticity being transferred into Polarization via 
an isentropic transition, respecting angular momentum conservation \cite{STAR:2017ckg}.
As well as a further confirmation of the fluid-like behavior of the medium, 
this observation opens the door to use polarization as a tool to study the medium's dynamics.   

\begin{figure}
    \centering
    \includegraphics[width=0.47\textwidth]{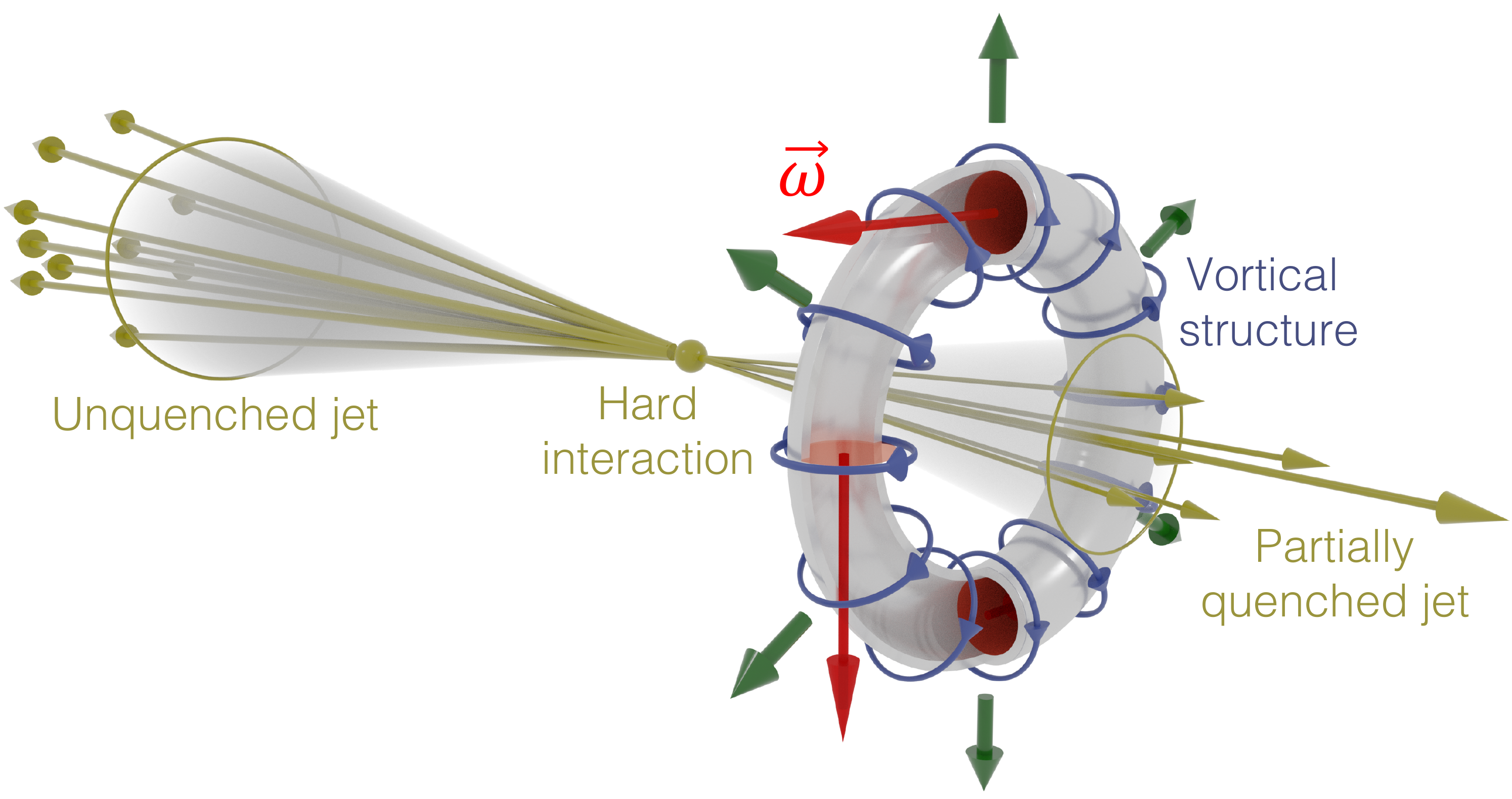}
    \caption{Schematic representation of the physical situation proposed.
    A hard parton generates a dijet structure and one of these jets is partially 
    quenched by the quark-gluon plasma, while the other is not. The quenched 
    portion of the jet introduces a momentum gradient in the fluid which in turn
    will generate a vortex ring.}
    \label{fig:SmokeRing}
\end{figure}

We propose to use polarization to understand the fate of
locally thermalized energy emitted by the jet. A schematic picture of 
the physical situation is shown in Fig.~\ref{fig:SmokeRing}. A hard parton
generates a dijet structure and one of these is partially quenched by the quark-gluon
plasma, while the other is not. The quenched portion of the jet introduces a 
initial velocity gradient in the fluid. As is known from everyday physics, 
smoke-rings, eddies and so on are ubiquitous in fluids when a velocity gradient is present.
This is certainly the case when a fast parton deposits energy into a medium. The 
only difficulty is, of course, that the jet's direction fluctuates event-by-event 
which vanishes after the event averaging.

This is, however, easily surmountable:  As argued in \cite{Betz:2007kg}, the interplay 
between vorticity and transverse expansion can be used to define a 
``jet production plane''. This insight can be sharpened into the definition of an 
experimental observable that ties the polarization direction, the angular momentum and 
a desired reference vector, which can be defined event-by-event. In this work, we shall 
focus on defining the reference vector as a high-$p_T$ trigger particle. This observable,
if measured to be non-zero in classes of events where jet suppression exists, 
would provide unique and compelling evidence that the
energy lost by the jet is indeed thermalized.
Moreover, it can be used to infer 
the medium's viscosity, provided the initial velocity gradients generated by the 
jet are quantified.

\section{A model for the jet thermalization}
\label{sec:IC}

Our first step is to choose a suitable model for the medium in which the jet will deposit 
(part of) its energy. We choose a model which incorporates three dimensional features, since
the $\Lambda$ polarization calculation we will perform later on will depend on the 
dynamics in all dimensions. The need to perform (3+1)D simulation imposes a heavy 
computational constraint. To make our work feasible, we take the average over a thousand 
initial conditions, generated with T$_{\rm R}$ENTo 3D \cite{Ke:2016jrd} configured for
simulations of \PbPb collisions at ${\sqrt{s_{NN}} = \SI{2.76}{\tera \electronvolt}}$,
all of them with impact parameter ${b = \SI{0}{\femto \meter}}$. The other parameters used to
generate these initial conditions were obtained from 
Ref.~\cite{Bernhard:2018hnz} (for parameters common to 2D and 3D \trento) and 
Ref.~\cite{Ke:2016jrd} (for parameters exclusive to 3D \trento). These 
are summarized in Table \ref{tab:trento_paramters}. All computations are made in a grid with spacing
equal to \SI{0.1}{\femto \meter} in the $x$ and $y$ directions\footnote{We attempted halving the 
grid spacing in $x$ and $y$ directions and our main results changed by only 1\%, at the expense of
a much greater computational effort.} and \num{0.2} in the spatial rapidity ($\eta_s$) direction.

\begin{table}[ht]
    \centering
    \caption{Input parameters for T$_{\rm R}$ENTo 3D.}
    \begin{tabular}{c c}
        \hline 
        Parameter                      & Value \\ \hline \hline
        Rapidity mean coefficient      & \num{0.0} \\
        Rapidity standard coefficient  & \num{2.9} \\
        Rapidity skewness coefficient  & \num{7.3} \\
        Skewness type                  & Relative skewness \\
        Jacobian                       & \num{0.75} \\
        Reduced thickness              & 0.007 \\
        Nucleon width                  & \SI{0.956}{\femto \meter} \\
        Nucleon minimum distance       & \SI{1.27}{\femto \meter} \\ \hline
    \end{tabular}
    
    \label{tab:trento_paramters}
\end{table}
We expect the event-averaged fluid background to give a good estimation on the polarization final observable. 
Karpenko and Becattini \cite{Karpenko:2018erl} showed that
the difference between event-by-event simulations and an averaged initial condition to be small, albeit
the source of $\Lambda$ polarization in their work is different from ours.

Now we turn our attention to the jet thermalization. We consider a scenario
of dijet creation inside the medium, where one jet will lose a 
negligible amount of energy and momentum while the other will be 
partially quenched, causing an asymmetry in jet emission. 
This is measured experimentally using the jet asymmetry observables $A_J$ and
$x_J$, defined as \cite{Aad:2010bu, Young:2011qx, Aaboud:2017eww, Sirunyan:2021jty}
\begin{align}
    x_J & \equiv p_{T_2}/p_{T_1}\,, 
    \label{eq:xJ}\\
    A_J & \equiv (E_{T_1} - E_{T_2})/(E_{T_2} + E_{T_1})\,.
    \label{eq:AJ}
\end{align}
The index ``1'' denote the trigger jet (the one that does not deposit 
energy and momentum in the medium) while the index ``2'' refers to 
the partially quenched jet.

From Eqs.~(\ref{eq:xJ}) and (\ref{eq:AJ}), one can obtain the momentum (energy)
of the quenched jet from the values of $x_J$ ($A_J$) and the
momentum (energy) of the trigger jet. Once $E_{T_2}$ and $p_{T_2}$ are 
determined, one may get the energy and momentum deposited in the medium 
as
\begin{align}
    \begin{aligned}
        p_{th} & = p_{T_1} - p_{T_2}\,, \\
        E_{th} & = E_{T_1} - E_{T_2}\,.
    \end{aligned}
\end{align}

We will use the data from \cite[Fig.~3]{Aad:2010bu} and 
\cite[Fig.~8]{Aaboud:2017eww} to determine the values of $p_{th}$ and $E_{th}$.
These are the distribution
of $d N/d A_J$ and $d N/d x_J$ for central \PbPb collisions at 
${\sqrt{s_{NN}} = \SI{2.76}{\tera \electronvolt}}$. The energy and 
momentum of the trigger jet in these measurements were 
${E_1 > \SI{100}{\giga \electronvolt}}$ and 
${p_{T_1} = \SI{89.5}{\giga \electronvolt \per \clight}}$. For the 
values of $A_J$ and $x_J$, we choose the ones that have the highest value
of multiplicity, i.e.~${A_J = \num{0.425}}$ and ${x_J = \num{0.525}}$. This gives
us ${E_{th} = \SI{59.6}{\giga \electronvolt}}$ 
and ${p_{th} = \SI{43}{\giga \electronvolt \per \clight}}$. 
This implies that the situation studied in what follows corresponds to a dijet
structure with a momentum of ${\SI{89.5}{\giga \electronvolt \per \clight}}$ 
for the unquenched jet and ${\SI{59.5}{\giga \electronvolt \per \clight}}$ 
for the partially quenched jet, noting that it is the latter
that defines the direction in which lambda polarization will be studied. 

\begin{figure*}[!htpb]
    \centering
    \includegraphics[width=.5\linewidth,trim=0 0cm 0cm 0cm,clip]{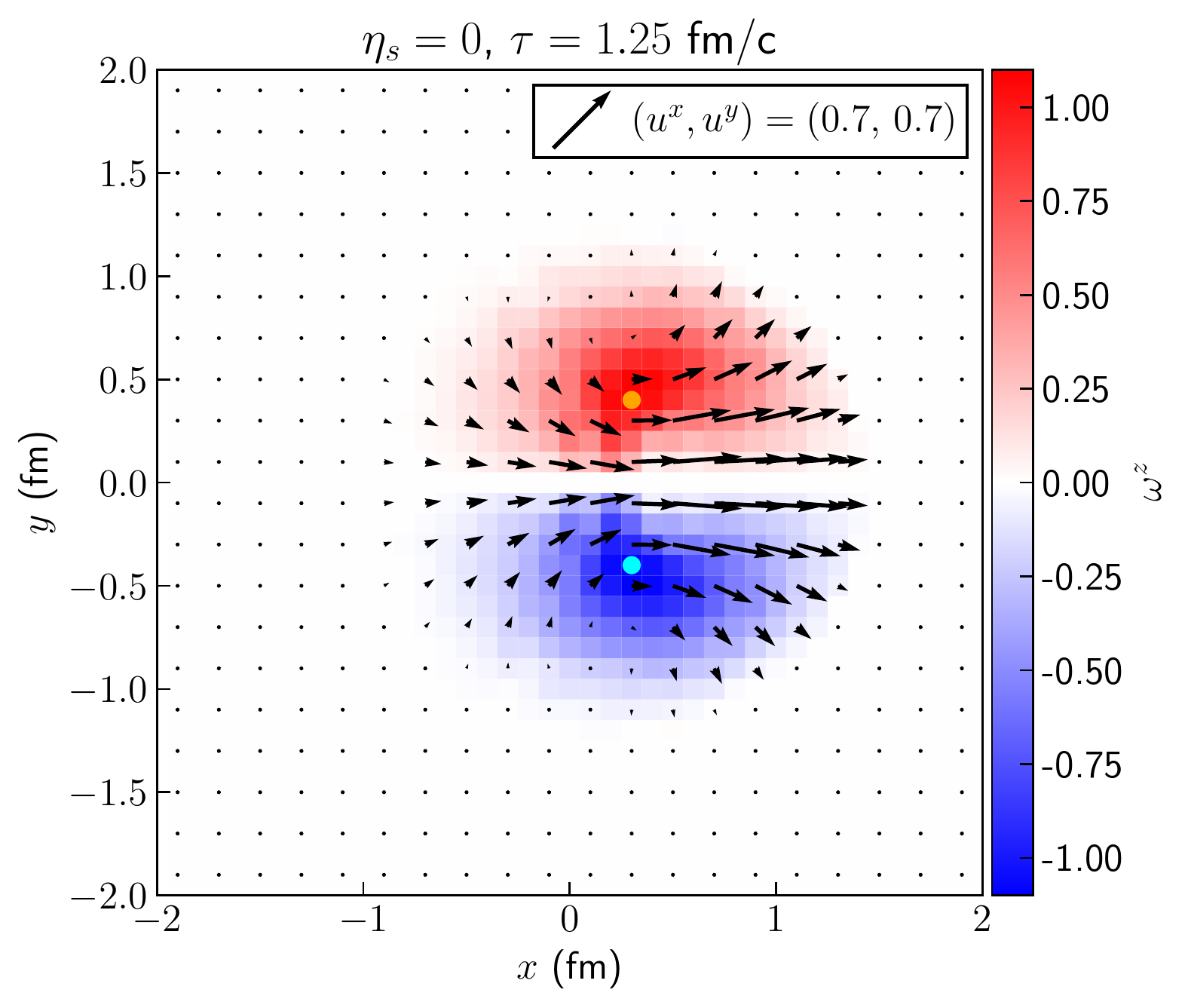}%
    \includegraphics[width=.5\linewidth,trim=0 0cm 0cm 0cm,clip]{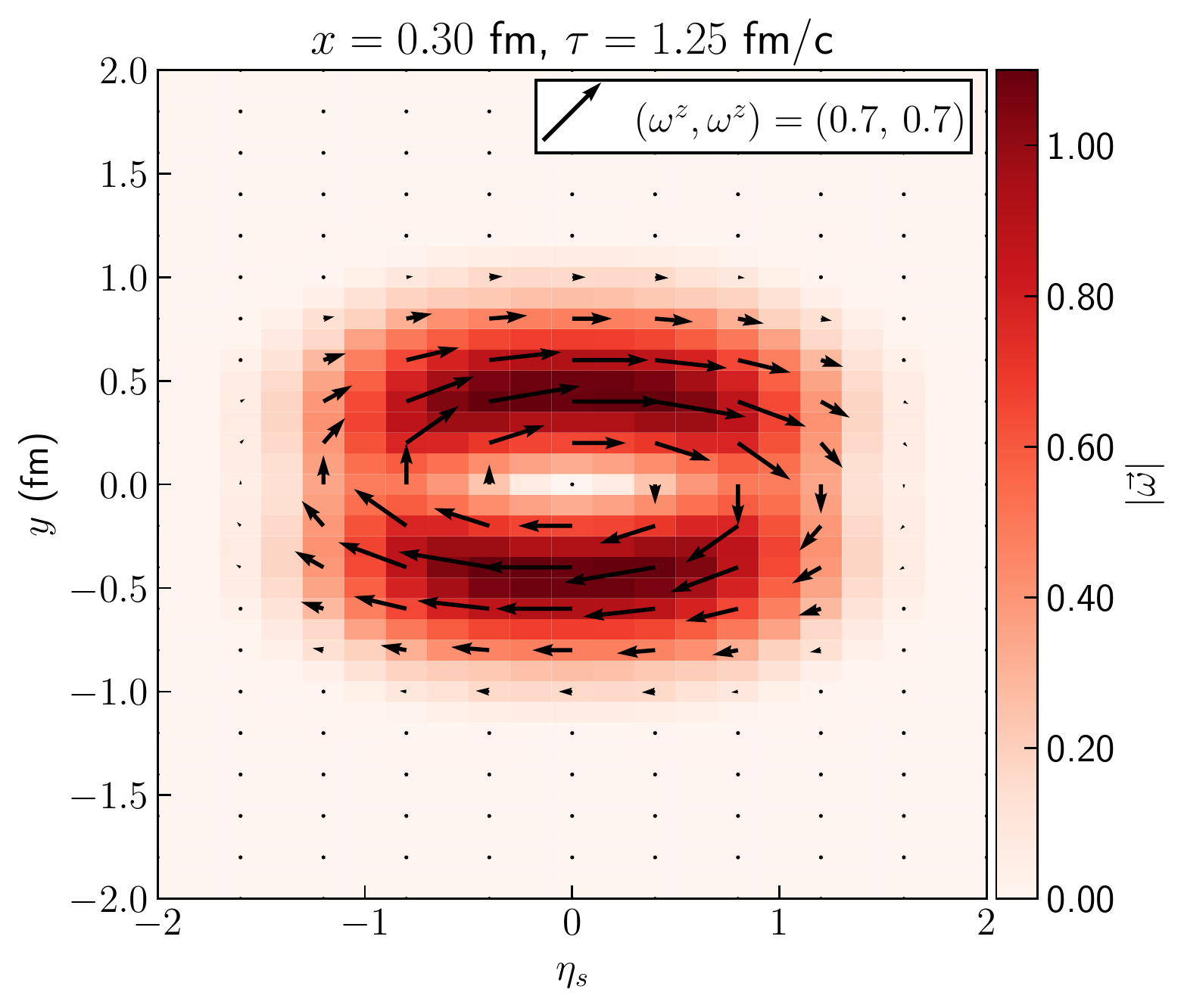}
    \caption{Vortex ring formed by the thermalized jet after 
             $\Delta \tau = \SI{1.00}{\femto \meter \per \clight}$ of
    hydrodynamic evolution. The jet deposited momentum in the $\hat{x}$ direction, i.e. to the
    right in the left panel and away from the viewer in the right panel.
    In the left panel, it is shown a slice of the system at $\eta_s = 0$. The color map shows 
    the $z$-component of vorticity vector defined in Eq.~(\ref{eq:vorticity_vector}). The arrows shows the $x$ and $y$ components of the 
    fluid's four-velocities. The dots marks the local maxima of 
    $|\omega^z|$. On the right panel, the system is sliced along the position
    $x=0.3$ fm.
    The color map shows $|\vec{\omega}|$ and the arrows shows the $y$ and $z$
    components of the vorticity vector.}
    \label{fig:ring_in_hydro}
\end{figure*}

The measurements that will be proposed later will be shown as a function 
of the difference between the azimuthal angle of the partially quenched jet and the 
emitted $\Lambda$. For simplicity, we choose the jet in the $x$-direction 
without loss of generality. With this choice, we may write the thermalized four-momentum 
as ${p^\mu_{th} = \begin{pmatrix} E_{th} & p_{th} & 0 & 0\end{pmatrix}}$ and build an energy-momentum tensor $T^{\mu \nu}$ following
\begin{align}
    T^{\mu \nu} = \frac{1}{V} \frac{p_{th}^\mu p_{th}^\nu}{E_{th}}\,,
    \label{eq:ExternalTmunu}
\end{align}
where $V$ is the volume over which the energy and momentum is deposited. The volume is chosen
to be an oblate spheroid centered on the origin of the system, with axis size equal to 
\SI{0.5}{\femto \meter} in the $x$ and $y$ directions and 
$\approx$ \SI{0.29}{\femto \meter} in the $z$-direction 
(which equates to $\eta_s \simeq 1$ at ${\tau = \SI{0.25}{\femto \meter \per \clight}}$).

We apply the Landau matching procedure $T^{\mu \nu}u_\nu = \varepsilon u^\mu$ to solve for the local energy density and flow velocity from the energy-momentum tensor in Eq. (\ref{eq:ExternalTmunu})
\begin{align}
        \varepsilon & = \frac{1}{V}\frac{E_{th}^2 - p_{th}^2}{E_{th}}\,,
        \label{eq:varepsilon}\\
        u^x & = \frac{p_{th}}{\sqrt{E_{th}^2 - p_{th}^2}}\,.
        \label{eq:ux}
\end{align}
The remaining spatial components of $u^\mu$ are zero and $u^\tau$ is obtained
by imposing the condition ${u^\mu u_\mu = 1}$. This procedure (energy-momentum tensor building 
and subsequent matching to a hydrodynamic-like energy-momentum tensor) was inspired by the 
procedure used for computing vorticity generated in the AMPT model in Ref.~\cite{Li:2017slc}.

By inserting in Eqs.~(\ref{eq:varepsilon}) and (\ref{eq:ux}) the values for
$E_{th}$ and $p_{th}$ obtained above, we obtain  $\varepsilon V = \SI{29}{\giga \electronvolt}$ and $v_x = \SI{0.69}{\clight}$, where
$V$ is the volume over which the energy density will be deposited. In our simulations, 
we rounded these values to $\varepsilon V = \SI{30}{\giga \electronvolt}$ and 
$v_x = \SI{0.7}{\clight}$. 
We verified that the injected energy-momentum generates on average 1\% more final state particles per unit of pseudo-rapidity.

\section{Fluid vorticity and polarization measurements}

\subsection{Jet induced fluid vorticity and $\Lambda$'s polarization}

The described initial condition is evolved with 3D viscous hydrodynamics
\cite{Schenke:2010nt, Schenke:2011bn, Paquet:2015lta}.
We use the lattice-QCD based equation of state from the HotQCD Collaboration \cite{Bazavov:2014pvz} and start the 
evolution at 
$\tau = \SI{.25}{\femto \meter \per \clight}$. The six independent components of the vorticity tensor are
then saved over a hypersurface of $T = \SI{151}{\mega \electronvolt}$. 
We then compute
the mean spin of $\Lambda$ following Eq.~(2) of Ref.~\cite{Becattini:2020ngo}, which we 
reproduce below for completeness.
\begin{align}
    \begin{aligned}
        P^\mu (p) &  = - \frac{1}{8m}\varepsilon^{\mu \rho \sigma \tau} p_\tau 
        \frac{\int d\Sigma_\lambda p^\lambda n_F (1-n_F)\omega_{\rho \sigma}}{\int d\Sigma_\lambda p^\lambda n_F}\,,\\
         n_F & = \frac{1}{1+\exp\left( \beta^\mu p_\mu - \mu Q/T \right)}\,, \\
         \omega^{\mu \nu} & =  -\frac{1}{2}\left(\partial^\mu \beta^\nu - \partial^\nu \beta^\mu\right) \quad \text{and}\quad \beta^\mu = \frac{u^\mu}{T}\,.
    \end{aligned}
    \label{eq:mean_polarization}
\end{align}
In our case, we do not consider baryon density and baryon currents and thus $\mu = \SI{0}{\mega \electronvolt}$.

With the six components of the vorticity tensor $\omega^{\mu \nu}$ we calculate a vorticity vector $\omega^\mu$
(inspired on the Pauli–Lubanski pseudovector), which will act as a proxy for the local spin polarization,
\begin{align}
\omega^\mu \equiv \varepsilon^{\mu \nu \rho \epsilon} u_\nu \omega_{\rho \epsilon}\,.
\label{eq:vorticity_vector}
\end{align}
In Figure~\ref{fig:ring_in_hydro}, we show the spatial distributions of $\omega^z$ (along a slice of $\eta_s = 0$) and $|\vec{\omega}|$ (along a slice of $x = 0.3$\,fm) at $\tau = \SI{1.25}{\femto \meter \per \clight}$. The external energy-momentum from the jet induces a ring-shaped concentration of vorticity around the jet axis during the hydrodynamic evolution.

\begin{figure}[h!]
    \centering
    \includegraphics[width=\linewidth]{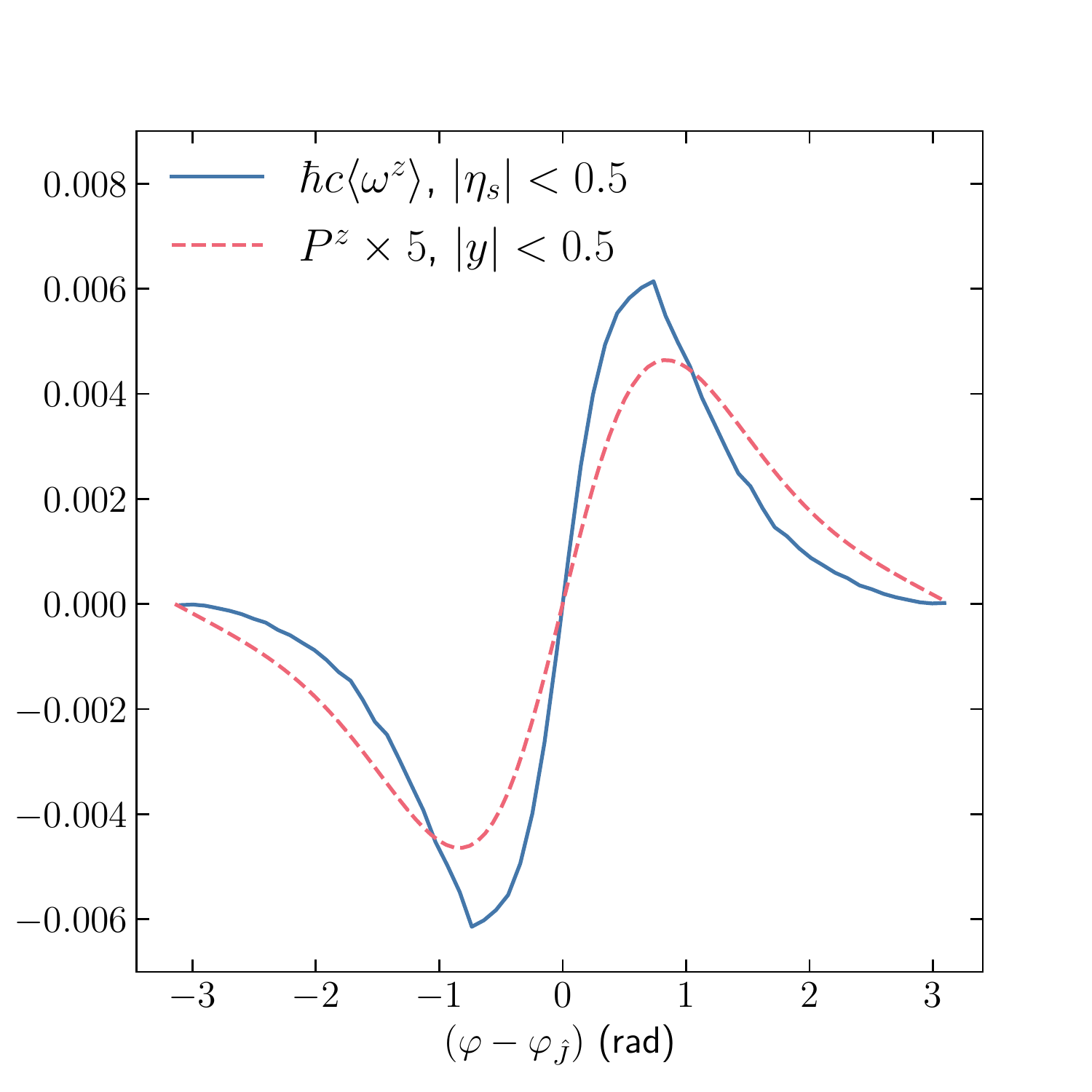}
    \caption{Comparison between the weighted average of the $z$-component of 
    the vorticity vector (see Eq.~\ref{eq:vorticity_vector}) and the weighted 
    average of the $z$-component of the $\Lambda$-polarization
    (see Eq.~\ref{eq:mean_polarization}) at mid-rapidity.}
    \label{fig:omega_time_hypersurface}
\end{figure}

To verify the vortical structures in the fluid velocity field are mapped to the spin polarization of emitted $\Lambda$,
we compare the averaged $\omega^z$ on the particlization hypersurface in the region $|\eta_s| < 0.5$ with the $\Lambda$'s $P^z$, averaged over the region $|y| < \num{0.5}$ and $p_T < \SI{3.0}{\giga \electronvolt \per \clight}$ in Fig.~\ref{fig:omega_time_hypersurface}.
To obtain the azimuthal angle of each cell on the particlization
hypersurface, we use the cell's four-velocity, i.e. $\varphi = \arctan(u^y/u^x)$.
Since the fluid is expanding in a mostly radial way, the velocity angle $\varphi$ is close to the spatial azimuthal angle of the cell.
Figure~\ref{fig:omega_time_hypersurface} shows that the sign of $\Lambda$ polarization correlates well with that of the fluid vorticity 
vector $\omega^\mu$ in Eq.~(\ref{eq:vorticity_vector}).

\begin{figure}[h!]
    \centering
    \includegraphics[width=\linewidth]{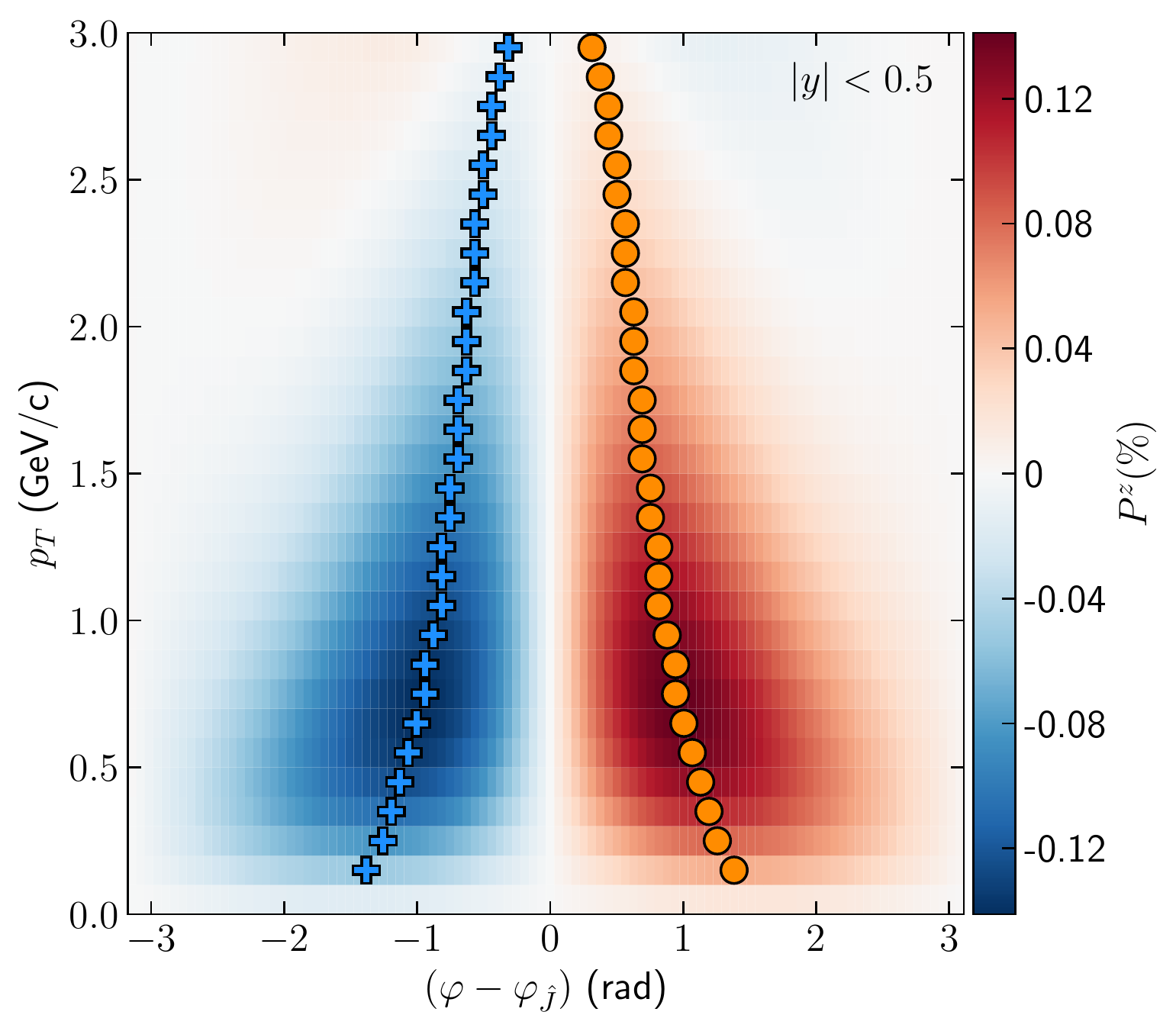}
    \caption{Distribution of the weighted average of the $z$-component of the polarization ($P^z$),
    using $\Lambda$-multiplicity as weight and as function of $p_T$ and the angular distance in the
    transverse place. The average considers only data in the range $|y| < \num{0.5}$.
    The orange/blue dots marks the bins where $|P^z|$ is highest for that $p_T$
    bin.}
    \label{fig:Pz_2D}
\end{figure}

Furthermore, we investigated the dependence of the $z$-component of the
$\Lambda$-polarization ($P^z$) with transverse momentum and the angular distance
(in the transverse plane) from the partially quenched jet, which we present in 
Fig.~\ref{fig:Pz_2D} as a color map.
The markers indicate the positions of the $|P^z|$'s maxima in each $p_T$-bin. 
The $|P^z|$'s maxima are closer to the jet axis at high $p_T$ than those at low $p_T$ bins.

\subsection{The ring observable}

We focused on the longitudinal component of polarization/vorticity for a jet that 
travels along the $+\hat{x}$ direction. Since the transverse
components are anti-symmetric with respect to rapidity/spatial-rapidity (see Fig.~\ref{fig:ring_in_hydro},
right panel), they will 
average to zero in the above calculations and we lose information about them. 
However, the formation of a vortex ring
due to our choice of initial condition has similarities with the vortex rings present
in p+A collisions which were studied in Ref.~\cite{Lisa:2021zkj}. There we
introduced the ring observable $\overline{\mathcal{R}}_{\Lambda}^{\hat{t}}$, which we
replicate below for completeness
\begin{align}
    \overline{\mathcal{R}}_{\Lambda}^{\hat{t}} \equiv
    \left\langle\frac{\vec{P}_{\Lambda}\cdot\left(\hat{t}\times\vec{p}_{\Lambda}\right)}{|\hat{t}\times\vec{p}_{\Lambda}|}\right\rangle_{p_T,\,y}\,.
    \label{eq:RLambda}
\end{align}
Here, ${\hat{t} = \hat{J}}$ is the axis direction of the jet\footnote{on our calculation, $\hat{J} = \hat{x}$}, 
and $\langle \cdot \rangle_{p_T,\,y}$ denotes an weighted average over transverse momentum 
(in the range $\SI{0.5}{\giga \electronvolt \per \clight} < p_T < \SI{3.0}{\giga \electronvolt \per \clight}$)
and rapidity (in the range $|y| < \num{0.5}$), using $\Lambda$ multiplicity as weight.
The use of $\overline{\mathcal{R}}_{\Lambda}^{\hat{J}}$ will filter most contributions to the polarization which were not
induced by the jet thermalization while allowing us to take into account effects in the
direction besides $\hat{z}$. We will focus on $\overline{\mathcal{R}}_{\Lambda}^{\hat{t}}$ from now on. 

\begin{figure}[h!]
    \centering
    \includegraphics[width=\linewidth]{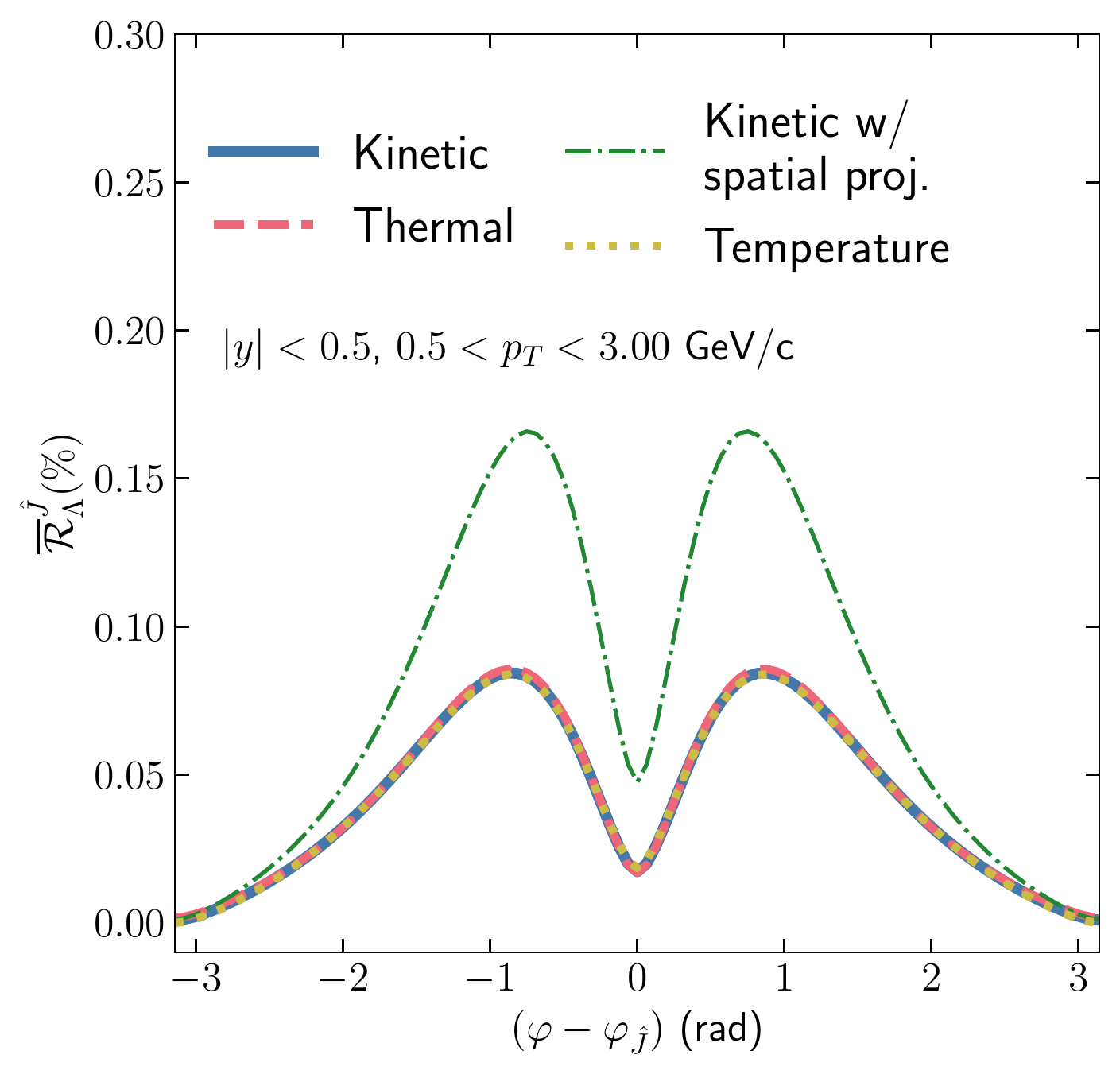}
    \caption{ $\overline{\mathcal{R}}_{\Lambda}^{\hat{t}}$ (see Eq.~\ref{eq:RLambda}) 
    computed from $\Lambda$-polarization calculations 
    using four types of vorticity tensor.}
    \label{fig:G1_vort_kind}
\end{figure}

The use of thermal vorticity, as shown in Eq.~\ref{eq:mean_polarization}, has been debated
in the literature~\cite{Becattini:2015ska, Ivanov:2018eej, Karpenko:2021wdm}. There are three other 
 definitions of vorticity which are popularly 
employed. The ``kinetic vorticity'' consists of the replacement $\beta^\mu \to u^\mu$
and is appealing because it can be more intuitively interpreted.
The ``temperature vorticity'' or ``T-vorticity'' relies on the replacement 
${\beta^\mu \to T u^\mu}$ and also allows vorticity 
generation by temperature gradients. Finally, there is the ``spatially projected kinetic
vorticity'' which replaces the derivative $\partial^\mu$ by 
${\nabla^\mu = (g^{\mu \nu} - u^\mu u^\nu)\partial_\nu}$. This has the effect of 
removing local acceleration terms from the kinectic vorticity. 
It also has a direct connection to the fluid vorticity in the 
non-relativistic limit. We show a comparison between 
the polarization results using these four different vorticity values in 
Fig.~\ref{fig:G1_vort_kind}. The fact that polarization from kinetic, thermal, and 
temperature vorticities are 
essentially equal implies that in this case the vorticity is predominately 
generated by gradients in velocity, not in temperature. The higher value for the 
polarization from the spatially projected kinetic vorticity implies that 
local acceleration (caused mostly by the fluid expansion) has the effect of
reducing the final $\Lambda$ polarization.

\begin{figure}[h!]
    \centering
    \includegraphics[width=\linewidth]{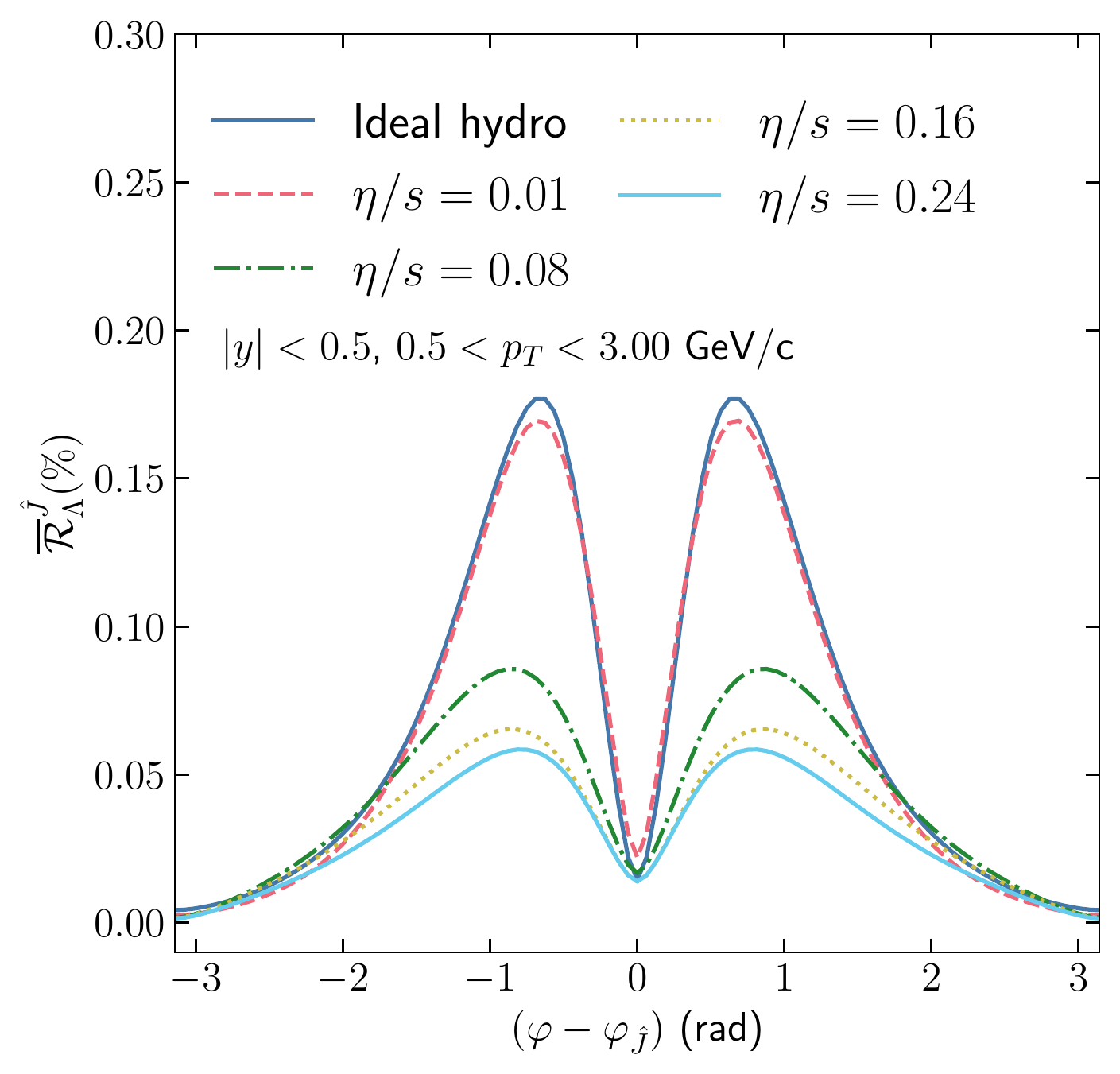}
    \caption{Distribution of $\overline{\mathcal{R}}_{\Lambda}^{\hat{J}}$ 
    (see Eq.~\ref{eq:RLambda}) for different specific shear viscosities.}
    \label{fig:vort_visc}
\end{figure}

We study the sensitivity of the ring observable $\overline{\mathcal{R}}_{\Lambda}^{\hat{t}}$ on medium's specific shear viscosity.
In addition to ${\eta/s = \num{0.08}}$, we perform calculations with $\eta/s = \num{0.00}$, $\num{0.01}$,
$\num{0.16}$ and $\num{0.24}$.
Figure~\ref{fig:vort_visc} shows that the medium's shear viscosity suppresses the ring observable $\overline{\mathcal{R}}_{\Lambda}^{\hat{t}}$\footnote{The angle where the signal is strong has a small 
dependence on viscosity as well.}. We observe a higher sensitivity of $\overline{\mathcal{R}}_{\Lambda}^{\hat{t}}$ to small viscosity values $\eta/s < 0.08$ than $\eta/s > 0.08$. 
This trend is consistent with the vorticity ring being quenched by the medium, an effect which will be stronger for higher viscosity, but that eventually gets saturated. This is in contrast to elliptic flow, which has a more or less uniform dependence with viscosity \cite{Romatschke:2007mq}.

\begin{figure}[h!]
    \centering
    \includegraphics[width=\linewidth]{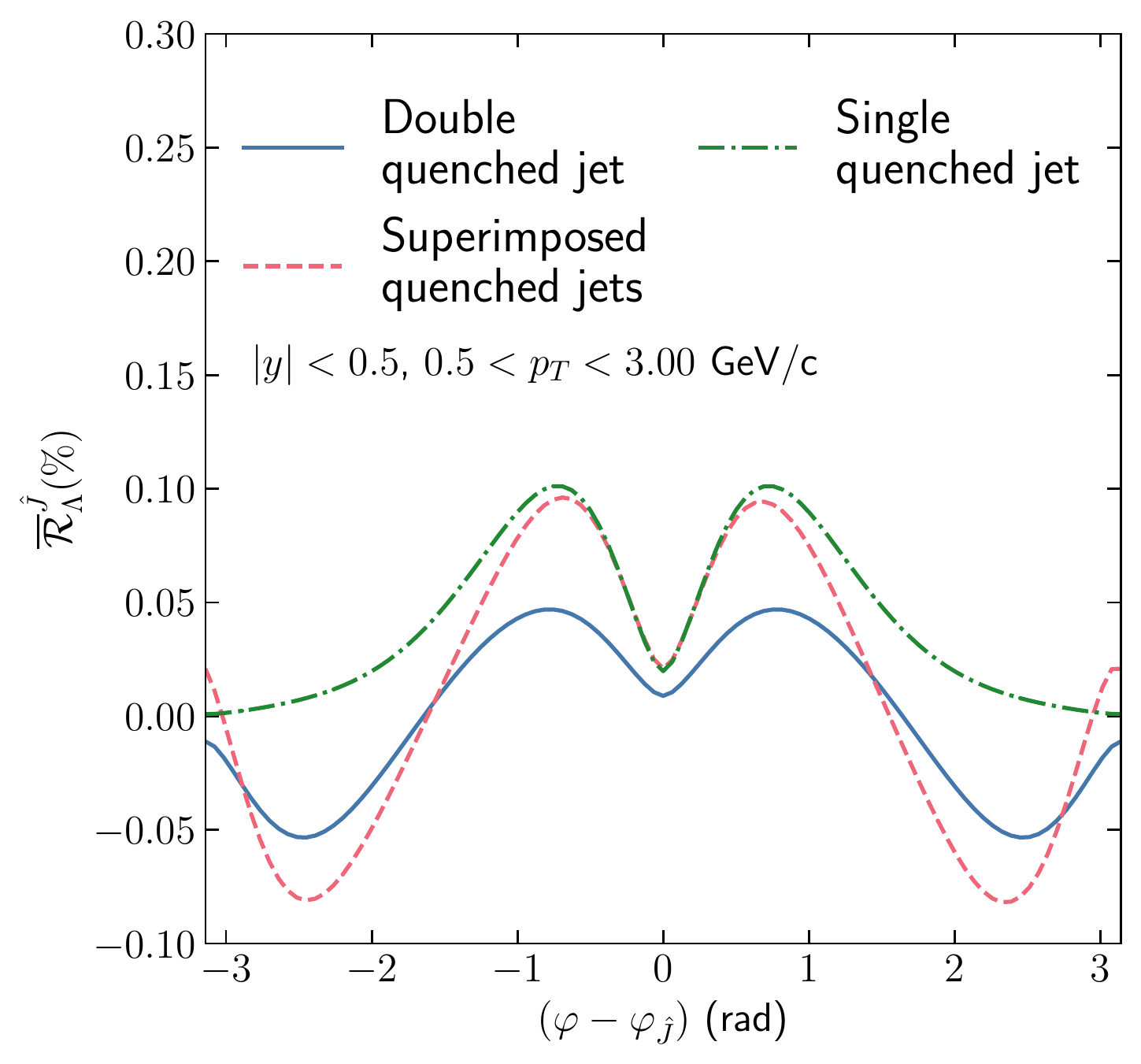}
    \caption{Comparison between the $\overline{\mathcal{R}}_{\Lambda}^{\hat{J}}$ in a 
    double-quenched jet scenario versus the single-quenched jet case. The blue curve 
    shows the result from the simulation and the red one by superimposing two 
    single-quenched jets (shown in green).}
    \label{fig:BtB_Jet}
\end{figure}

It is possible to argue that a jet which is quenched at the center
of the system will not be accompanied by an unquenched jet. Instead, there 
would be a pair of quenched jets, inducing a pair back-to-back vortex rings. One could 
approximately treat the medium excitation from the two quenched jets as independent superposition
(after rotating one of them by $\SI{\pi}{\radian}$). However, this would 
neglect the possibility of interactions between 
the two vortexes during the hydrodynamic evolution. We investigate the possibility of
a double-quenched jet by displacing the energy-momentum deposition to 
$x = \SI{.6}{\femto \meter}$. In the sequence, we add a second one at 
$x = \SI{-.6}{\femto \meter}$ with momentum in the opposite direction of the 
first. We compare the superposition scenario with the full simulation in Fig.~\ref{fig:BtB_Jet}.
It is clear to see that 
the superposition scenario has a polarization which is 
almost double the one where we evolve the two quenched jets, indicating
the interaction between them during hydrodynamic evolution is crucial and has a self-canceling effect.

\section{Conclusions}

We modeled the thermalization of the energy-momentum from a hard parton as a ``hot spot''
which propagates inside fluid dynamic simulations.
Such configuration of velocities will generate 
a vortex ring, which can be quantified by the vorticity of the fluid. The 
vorticity will lead to the emission of polarized hadrons on the particlization
hypersurface as described in \cite{Becattini:2015ska,Becattini:2020ngo}.

To obtain the energy and momentum deposited in the medium by the jet 
thermalization, we assumed a jet with a
transverse momentum of $\SI{89.5}{\giga \electronvolt \per \clight}$ that would 
deposit approximately 40\% of its 
energy in the medium, motivated by \cite[Fig.~3]{Aad:2010bu} and \cite[Fig.~8]{Aaboud:2017eww}. 
The polarized hadron emission would accompany a partially quenched jet, meaning that 
experimentally any analysis aiming to measure this effect 
would have to focus on an asymmetric jet pair, with the higher momentum jet having momentum 
of the order of $\SI{90}{\giga \electronvolt \per \clight}$ and the lower momentum being of 
order $\SI{60}{\giga \electronvolt \per \clight}$. 
Other options, such as using high-momentum trigger particles, will also be investigated
in future work. 

We computed the polarization of the $\Lambda$
hyperon due to the vorticity caused by our model of jet thermalization. We showed
that, for this specific case, the effects are dominated by velocity gradients and 
thus there is little difference in using thermal vorticity versus other
definitions which are often suggested in the literature. We also showed that 
the strength of the signal is highly sensitive to
the fluid's shear viscosity.

The angular distribution of the ring observable $\overline{\mathcal{R}}_{\Lambda}^{\hat{t}}$
in the transverse plane with respect to the 
quenched jet peaks in the range \SIrange{0.5}{1.0}{\radian}, depending on transverse
momentum. This position depends also on the shear viscosity as well, albeit
in a more subtle way than the polarization amount. We also showed that the 
addition of a second quenched jet will not significantly affect the region
where $\overline{\mathcal{R}}_{\Lambda}^{\hat{J}}$ peaks. Instead, it will
only dampen the overall magnitude in addition of an expected additional lobe 
in the opposite direction.

We point out that, despite the effect being of the order of only a few tenth 
of a percentiles, the proposed ring observable $\overline{\mathcal{R}}_{\Lambda}^{\hat{t}}$ should be measurable by experiments, since it has the same 
of magnitude as reported per ALICE and STAR for the global $\Lambda$-polarization 
\cite{STAR:2017ckg, Acharya:2019ryw}.
We also inspected the typical maximum value found for $\overline{\mathcal{R}}_{\Lambda}^{\hat{J}}$.
We found that $\overline{\mathcal{R}}_{\Lambda}^{\hat{J}} < 0.25 \%$ always, peaking
in the $p_T$ range of $\SI{.5}{\giga \electronvolt \per \clight} < p_T < \SI{1.0}{\giga \electronvolt \per \clight}$.

We devote a future study to quantify the effects of event-by-event fluctuations in the fluid on $\overline{\mathcal{R}}_{\Lambda}^{\hat{t}}$.

We note that the discussed jet induced polarization effect requires both color opacity and rapid thermalization. Thus, it is very likely present
in AA and might disappear in pp and pA collisions (which may have rapid thermalization,
but very small opacity). Since the reference is 
a high momentum trigger rather than a global quantity like the reaction plane, 
it should be possible for experiments to examine events with one 
$\Lambda$ and one high momentum triggered hadron to verify this effect. If it
turns out that indeed $\overline{\mathcal{R}}_{\Lambda}^{\hat{J}}$ is non-zero for $AA$ events, one could proceed
to do more detailed model-data comparisons as a way to constrain viscosity and jet energy loss.

\section*{Acknowledgments}

WMS, JGPB, DDC, JT and GT are supported by FAPESP projects 17/05685-2 (all), 19/05700-7 (WMS), 19/16293-3 (JPB) and 17/06508-7 (GT).
MAL is supported by the U.S. Department of Energy grant DE-SC0020651 and acknowledges support of the Fulbright Commission of Brazil.
CS is supported by the U.S. Department of Energy under grant number DE-SC0013460 and the National Science Foundation under grant number PHY-2012922.
GT acknowledges CNPQ bolsa de produtividade 301432/2017-1.

\bibliographystyle{elsarticle-num}
\bibliography{references}
\end{document}